\documentclass[twoside,fleqn]{article}
\usepackage{espcrc2}
\usepackage{graphicx}
\usepackage[figuresright]{rotating}

\def    \nn             {\nonumber}
\def    \=              {\;=\;}
\def    \frac           #1#2{{#1 \over #2}}

\def    \lsim           {\raisebox{-3pt}{$\>\stackrel{<}{\scriptstyle\sim}\>$}}
\def    \gsim           {\raisebox{-3pt}{$\>\stackrel{>}{\scriptstyle\sim}\>$}}
\def    \esim         {\raisebox{-3pt}{$\>\stackrel{\sim}{\scriptstyle{-}}\>$}}
   
\newcommand     \sss            {\scriptscriptstyle}

\newcommand     \avg[1]         {\left\langle #1 \right\rangle}

\newcommand     \ba             {\begin{eqnarray}}
\newcommand     \ea             {\end{eqnarray}}
\newcommand     \be             {\begin{equation}}
\newcommand     \ee             {\end{equation}}

\newcommand     \epem           {\ifmmode{e^+e^-}\else{$e^+e^-$}\fi}
\newcommand     \lambdamsb     {\ifmmode
          \Lambda_4^{\rm \scriptscriptstyle \overline{MS}} \else
         $\Lambda_4^{\rm \scriptscriptstyle \overline{MS}}$ \fi}
\newcommand     \MSB            {\ifmmode {\overline{\rm MS}} \else 
                                 $\overline{\rm MS}$  \fi}

\newcommand     \ptmin     {\ifmmode p_{\scriptscriptstyle T}^{\sss min} \else
                           $p_{\scriptscriptstyle T}^{\sss min}$ \fi}

\newcommand\as{\alpha_{\sss S}}

\def\vcb{V_{cb}}
\def\vub{V_{ub}}
\def\vtd{V_{td}}
\def\vts{V_{ts}}
\newcommand\epspoeps{\ifmmode \frac{\epsilon'}{\epsilon}
              \else   $\epsilon'/\epsilon$ \fi} 

\def    \calO   {{\cal O}}

\def \pt   {\mbox{$p_{\scriptscriptstyle T}$}}                
\def \et   {\mbox{$E_{\scriptscriptstyle T}$}}

\def \to   {\mbox{$\rightarrow$}}

\newcommand \vckm{\ifmmode{V_{\rm CKM}
    }\else{$V_{\rm CKM}$}\fi} 
\newcommand \jpsi{\ifmmode{J/\psi
    }\else{$J/\psi$}\fi} 
\newcommand \cpviol{\ifmmode{\not
    \!\!\!{\rm CP}}\else{$\not \!\!\!{\rm CP}$}\fi}
\newcommand{\vubovcb}{\left | \frac{V_{ub}}{V_{cb}} \right |}

\def\gappeq{\mathrel{\rlap {\raise.5ex\hbox{$>$}}
{\lower.5ex\hbox{$\sim$}}}}
\def\lappeq{\mathrel{\rlap{\raise.5ex\hbox{$<$}}
{\lower.5ex\hbox{$\sim$}}}}
\title{\flushright{\normalsize hep-lat/9910019}\\
\flushleft{Results and Perspectives in HEP, vis-a-vis Lattice QCD}}
\author{M.L. Mangano\address{CERN, Theoretical Phisics Division 1211
    Geneva 23, Switzerland} \thanks{This work was supported in part by
    the EU Fourth Framework Programme ``Training and Mobility of
    Researchers'', Network ``Quantum Chromodynamics and the Deep
    Structure of Elementary Particles'', contract FMRX--CT98--0194 (DG
    12 -- MIHT).}  }

\begin{document}
\begin{abstract}
I review in this presentation some aspects of phenomenology in High
Energy Physics which are related to recent and possibly future
progress in lattice QCD. In particular, I cover (i) the extraction of CKM
matrix elements from $B$ physics, (ii) the determination of \epspoeps,
as well as (iii) some issues emerged in the physics of high energy
jets produced in hadronic collisions, where input from
non-perturbative calculations would benefit our capability to perform
better theoretical predictions.
\end{abstract}

\maketitle

\section{INTRODUCTION}
The last couple of years have given us an impressive series of
important new results in HEP. Most of them, if not all, are in the
field of flavour physics. We obtained conclusive evidence for neutrino
mixing and therefore for neutrino masses (SuperKamiokande); new
results on $\epsilon'/\epsilon$ (KTeV, NA48); the first strong
evidence for CP violation in the $B$ system (CDF); the first direct
limits on $B_s$ mixing (LEP, SLC).  Impressive progress is thus taking
place in the area of flavour physics, and these results are just the
appetizer for much more to come in the next few years. New data are in
fact soon expected from the B-factories (BaBar and Belle, both of
which successfully started in the Summer their operations, and CLEO),
from the upcoming run of the Tevatron (CDF, D0), from the analyses of
the full datasets of KTeV and NA48, from the first operation of the KLOE
experiment at the Daphne $\phi$ factory, and from the new
long-baseline neutrino beam from KEK to SuperKamiokande. Finally, a
series of new experiments has recently been (or hopefully will soon be)
approved: searches for rare kaon decays (FNAL, BNL), searches for
$\mu\to e\gamma$ decays (KEK, PSI) and new long baseline
neutrino-oscillation experiments (MINOS at FNAL, NGS at CERN/Gran
Sasso).

On the theoretical side, the problem of flavour is among the most
interesting ones. Flavour physics provides a key peephole to explore
and possibly uncover the existence of new phenomena beyond the
Standard Model.  In supersymmetry, the problem of flavour is receiving
nowadays as much attention as the problem of SUSY breaking, if not
more, and the two problems are universally believed to be intimately
related.  The observation of flavour phenomena not explained within
the SM would provide valuable input for the understanding of physics at
the SUSY breaking scale.

For the experimental information to be useful in this direction, it is
of fundamental importance that accurate data be compared with accurate
SM predictions.  With the exception of the observation of neutrino
masses, the interpretation of all the new input on flavour physics
coming from the observations listed above heavily
relies on non-perturbative physics, which lattice calculations are
best suited to carry out.  Almost anything being calculated on the
lattice is fundamental to interpret the available and forthcoming
data: the efforts of the lattice community are therefore much
appreciated by all of us living in the continuum!

I will review in this talk the status of the experimental inputs, and
the perspectives for future improvements. As an outsider in the field
of lattice, I hope this review will be useful to those of you mostly
involved in the numerical aspects of lattice physics. I apologise to
those of you who are experts in the phenomenological aspects I will
discuss, and who may find this review incomplete, or too
naive. Hopefully the bibliography will help filling in some gaps.  I
will concentrate on the following selection of topics: the status of
the extraction of the CKM matrix elements from $B$ physics, and the
measurement of $\epsilon'/\epsilon$. In both cases, I will make heavy
use for this presentation of very good, recent extensive reviews of
the subjects.  In addition, I will cover a couple of topics taken from
current high-energy phenomenology: prompt-photon production in fixed
target, and production of high-energy jets in hadronic collisions.
Evidence is emerging that a better understanding of the interface
between the perturbative and non-perturbative domain is required to
solve some problems in this area. Whether the lattice approach can
provide some useful input to address these issues, it is too early to
say, and hard for me to judge. I hope the following discussion will
anyway serve as a stimulus for some of you to try address these
questions.

\section{CKM PARAMETERS FROM B PHYSICS}
\vckm\ is the matrix describing the couplings of the weak charged
currents of quarks to the $W$ gauge boson:
\be L_{EWK} \sim V_{ij} \; \overline{Q}^i \,\gamma_\mu(1-\gamma_5)\,Q^j
\; W^{\mu}
\ee
With 3 generations, the CKM matrix can be parameterised as follows:
\be
\left( \begin{array}{ccc}
    V_{ud} & V_{us} &  { V_{ub}} \\V_{cd} &  V_{cs} & { V_{cb}} \\
   { V_{td}} & 
   {  V_{ts}} & { V_{tb}}
\end{array}
\right) =
\ee
\be
\left( \begin{array}{ccc}
    1 - \frac{\lambda^2}{2}            &   \lambda   &
    { A\lambda^3(\rho-i\eta)}
\\
     - \lambda         &   1 - \frac{\lambda^2}{2}         &
    { A\lambda^2}
\\
    { A\lambda^3(1-\rho-i\eta)}  & { -A\lambda^2}  & 1
\end{array}
\right) 
\ee
Here $\lambda=\sin\theta_c$ is the sine of the Cabibbo angle,
and we neglected  ${\cal O}(\lambda^4)$
corrections. In addition to the Cabibbo angle, 3 more parameters are
required to parametrize the matrix: 2 real numbers, and a phase. 
A direct determination of the \vckm\ elements~\cite{PDG} should be
obtained from  tree-level processes, to avoid possible
contaminations from  physics beyond the SM.
%\begin{itemize}
%\item[$\vert V_{ud}\vert$:] $0.9740\pm 0.0010$ from ``$n\to p e
%\bar\nu$'' \cite{PDG}.
%\item[$\vert V_{us}\vert$:] $0.2196\pm 0.0023$ from $K\to \pi \ell
%\bar\nu$  \cite{PDG}.
%\item[$\vert V_{cd}\vert$:] $0.224 \pm 0.016 $ from $\nu N \to \mu c $
%\cite{PDG}.
%\item[$\vert V_{cs}\vert$:] $1.04 \pm 0.16 $ from $D\to \bar{K} e^+
%  \nu$ \cite{PDG}.
%\item[$\vert V_{cb}\vert$,] from $b\to c$ decays (see later for details).
%\item[$\vert V_{ub}\vert$,] from $b\to u$ decays (see later for details).
%\end{itemize}
Unfortunately, this is not possible for all entries: no direct
measurement of $ V_{td,ts}$ has been possible so far, and a crude
measurement of $V_{tb}\sim 1$ is only possible if one assumes
knowledge of the $t\bar t$ production cross section at the Tevatron.
The comparison with observables induced by higher-order processes
(e.g. $K^0-\bar{K}^0$ mixing, $\propto (V_{cs} V_{cd}^*)^2$), provides
consistency checks, potential measurements of relative phases, and 
possible information on BSM phenomena.

\subsection{$\bf \vert V_{cb} \vert$}
$\vert V_{cb} \vert$ is measured from both exclusive and inclusive
semileptonic decays of $B$ mesons with charmed hadrons in the final
state. Since these processes proceed at
tree level, the determination of $\vert V_{cb} \vert$ is with very
good accuracy free of possible contaminations from new physics.

The inclusive extraction is obtained from the measurement of the
inclusive lifetime, and of the semileptonic branching ratio (B).
% error estimates from Uraltsev)}:
The theoretical estimates of lifetime (for reviews, see e.g.
ref.~\cite{Neubert:1996qg})
have uncertainties from the
knowledge of the $b$ and $c$ quark masses, of the heavy quark kinetic
energy inside the hadron, and of higher-order corrections to the
perturbative expansions in powers of $\as$ (known to NNLO), and in
powers of $1/m_Q$ (known to ${\cal  O}(1/m_Q^2)$). 
The contribution of 
these sources of uncertainties to the determination of $\vert V_{cb} \vert$
is given by the following formula (uncertainties form the recent
compilation in ref.~\cite{LEPSC}):
\ba 
\vert V_{cb} \vert \= 0.0411 \sqrt{\frac{B(B\to\ell\nu X_c)}{0.105}}
\sqrt{\frac{1.55{\rm ps}}{\tau(B)}}\times && \nn \\
 \left[1-0.024\frac{\mu_\pi^2-0.5}{0.1\,{\rm GeV}^2}\right] \times && \nn \\
 \left[ 1 \pm0.030_{\vert PT}
            \pm0.020_{\vert \Delta m_b}
            \pm0.024_{\vert 1/m_Q^3} \right] &&
\ea
%where $\Delta m_b=60{\rm MeV}$. 
The term proportional to
$\mu_{\pi}^2$ parametrizes the quark kinetic energy, and
reflects the uncertainty in the extraction of the $b-c$ mass
difference from the HQET relation~\cite{Neubert:1996qg,LEPSC}:
\ba m_b-m_c &=& \avg{M_B}-\avg{M_D}+
\mu_\pi^2\left(\frac{1}{2m_c}-\frac{1}{2m_b}\right) \nn \\
  &+&{\cal
  O}(1/m^2_{c,b}) \nn \\ 
  &\sim&
 3.50~{\rm GeV}+40~{\rm MeV} \cdot\frac{\mu_\pi^2-0.5}{0.1\,{\rm GeV}^2}
\ea
The latest results on the inclusive extraction of $\vert \vcb \vert$
come from the average of the LEP measurements~\cite{LEPVcb}:
\be
        \vert V_{cb} \vert \= (40.75
         \pm 0.41_{exp} \pm 2.04_{th} ) \times 10^{-3}
\ee
Notice that the theoretical error is much larger than the current
experimental uncertainties.

The exclusive extraction requires the theoretical knowledge of the form
factor for $B\to D^* \ell \nu$ at zero recoil. 
Spin symmetry in the $m\to \infty$ limit gives
$F(1)=1$~\cite{Neubert:1996qg}. Finite mass corrections are evaluated
within HQET, giving (for a recent review see~\cite{Bigi:1999dv}; this
value was taken from the compilation in ref.~\cite{LEPSC})
$F_{D^*}(1)=0.88\pm0.08$, which is consistent with the most recent
determinations from the lattice presented at this
Conference~\cite{Hashimoto:1999yp}, $F_{D^*}(1)=0.935\pm0.035$.
%At this time {\tiny (Neubert, $D^*\ell\nu$)}:
%\[
%\magenta
%\vert V_{cb}^{excl}\vert \times 10^{4} =  388 \pm 21_{exp} \pm 13_{th}
%\]
%Used with an average LEP:
%\[ F(1)\vert V_{cb}\vert = (33.8\pm2.2)\cdot 10^{-3}\]
%gives:
%\[\red
The current average of the exclusive LEP results gives~\cite{LEPVcb}:
\be
  \vert V_{cb}^{excl}
        \vert = (38.4\pm2.5_{\rm exp}\pm2.2_{\rm th})\cdot 10^{-3}
\ee
% Winter 99 BR(b->ell X) form LEP: .1078\pm 0.0021, from LEPHF/99-01
and the  world average, including CLEO, is~\cite{Artuso}:
\be
\vert V_{cb}^{excl} \vert \times 10^{4} = 385 \pm 9_{st} \pm 16_{syst}
        \pm 26_{th}\; ,
\ee
which includes the correlations of experimental and theoretical
uncertainties of the  various experimental results.

The current world average of the CLEO and LEP 
inclusive and exclusive results~\cite{Artuso} is:
\be
\vert V_{cb}\vert \times 10^{4} = 400 \pm 4_{exp}\pm  21_{th}\; .
\ee
Again, notice the large gap in accuracy between the experimental
and the theoretical precisions.
%At this time
%(LEP/CLEO$II$ average by Parodi, Roudeau, Stocchi 1999)
%\be
%\vert V_{cb}\vert \times 10^{4} = 388 \pm 21~({\rm excl}) \; ,
%410 \pm 15~({\rm incl})
%\ee
%Proposed for the Summer '99 LEP averages:
%\ba
%\vert V_{cb} \vert &=& 0.0XXX^{(*)} \sqrt{\frac{B(B\to\ell\nu X)}{0.108}}
%\sqrt{\frac{1.55{\rm ps}}{\tau(B)}} \nn \\
%   &\times& (1\pm 0.05)
%\ea

%Notice however:\\[0.5cm]
%\epsfig{figure=vcbmoments.eps,height=0.55\textheight,clip=}

\subsection{$\bf \vert V_{ub}\vert $}
The measurement of $\vert V_{ub}/V_{cb}\vert = \vert\lambda (\rho
-i\eta)\vert$ provides the first direct constraint on the
$(\rho,\eta)$ plane, in the form of a circle centered
around the origin.
%\centerline{                                                      
%\epsfig{figure=vub.eps,height=0.23\textheight,clip=}}
$\vert V_{ub} \vert$ is  measured in charm-less $B$ decays. In
particular, the used observables are:
\begin{itemize}
  \item exclusive decays such as $B\to \pi \ell \nu$, $B\to \rho \ell \nu$;
  \item the end-point spectrum in inclusive $B\to \ell X$ decays;
  \item low-mass hadronic recoils  $X_u$ in $B\to \ell X_u$
\end{itemize}
Each one of these tecniques has different sets of theoretical and
experimental systematics, which limit the potential accuracy of the
measurement. For example, measurements based on the study of the
end-point lepton spectrum rely on the assumed knowledge of the lepton
spectrum away fom the end-point, in the low-energy region where
production is dominated by charmed decays. The measurement done
requiring a low-mass hadronic recoil system can be affected by the
assumptions made on the exclusive structure of the final state, since
these enter in the experimental definition of the sample itself.

Exclusive
decays are experimentally much cleaner, but the theoretical
interpretation is based on the assumed knowledge of the exclusive form
factors for heavy-to-light
transitions. The application of HQET is not reliable, since the $b$
decays to a light quark. One therefore has to rely on phenomenological
models, or on lattice calculations. These,  however,
are less trustable for $B\to
\pi\ell\nu$  than for $B\to
\rho\ell\nu$ transitions. Current error estimates on the exclusive
form factors are of the
order of $\pm 15\%$. Some new results have been presented at this
conference, and were reviewed in the plenary talk by
Hashimoto~\cite{Hashimoto:1999bk}. 

The latest measurements from CLEO~\cite{Behrens:1999vv} rely on $B\to
\rho\ell\nu$ exclusive decays, and the result is:
\be
  10^{3}\cdot \vert V_{ub} \vert = 3.25\pm 0.14_{stat}
  {+0.21\atop-0.29}_{syst} \pm 0.55_{th}
\ee
%\item $\vert V_{ub} \vert/\vert V_{cb} \vert = 0.080\pm 0.017$ (CLEO
%  inclusive) \\
%$\vert V_{ub} \vert/\vert V_{cb} \vert = 0.104\pm 0.019$ (LEP
%  inclusive) 
%(S.Stone 1999):\\
%\hfill \epsfig{figure=vub_cb.eps,width=0.9\textwidth,clip=}

In the case of inclusive decays, calculations rely on the OPE and HQET.
Theoretically, the main uncertainties come from
higher-order PT corrections, and from $m_b$, since the phase-space is
proportional to $m_b^5$. The relation between $\vert V_{ub} \vert$
and  $B(B\to\ell\nu X_u)$, which is currently used by the LEP experiments, is
given by the following expression~\cite{Bigi:1997dn}
\ba \vert V_{ub}^{incl} \vert &=&
  0.00445 \sqrt{\frac{B(B\to\ell\nu X_u)}{0.002}} \sqrt{\frac{1.55{\rm
      ps}}{\tau(B)}}\times \nn \\ 
&& \!\!\!\! (1\pm 0.020_{PT}\pm 0.035_{\delta
  m_b=60~{\rm MeV}})  \label{eq:vub}
\ea 
The 60~MeV uncertainty on $m_b$ ($m_b(1\mbox{GeV})=4.58\pm0.06$), although
perhaps a bit optimistic, is consistent with the most recent estimates
based on Sum Rules and NNLO QCD studies of the $\Upsilon$
spectrum\cite{Melnikov:1998ug}, as well as with the most recent
unquenched lattice estimates \cite{Gimenez:1999en}.

It should be pointed out that even if it were possible to calculate
with high accuracy the full, charm-less inclusive rate, large
uncertainties would still remain in the extraction of $B(B\to\ell\nu
X_u)$  from the data. This is 
because the phase-space region used by experiments is only a very
small fraction of the total one, in order to suppress the large
$b\to c X$ backgrounds.
For this reason, the measurement of $B(B\to\ell\nu
X_u)$ is also affected by uncertainties in the modeling of the
structure of the final states, which are in
principle of theoretical origin. The most recent LEP results
give~\cite{LEPVub}:
\ba
&&10^3 \times  B(B\to\ell\nu X_u)= \nn \\
&&1.67 \pm 0.35_{exp}\pm 0.38_{b\to c} \pm 0.20_{b\to u}
\label{eq:btou}
\ea
where the last two errors come from the modeling systematics.
Using this result and eq.~(\ref{eq:vub}), LEP obtains:
\be
\vert \vub \vert = (4.05 {+0.62 \atop -0.74}) \; 10^{-3}
\ee
where the overall error is approximately 40\%  experimental and
60\% theoretical. 

%Recent results from LEP are based on the ??, and give:
%\epsfig{figure=lepbtou.eps,height=0.6\textheight,clip=}\\
%\ba 
%10^3\times\vert \vub \vert &=& 4.21 \; 
% \pm0.02_{\tau}\pm0.16_{HQE}
%\nn \\
%&& 
%\overbrace{
%{\sst +0.42\atop \sst -0.46}_{exp}\; 
%{\sst +0.46\atop \sst -0.52}_{b\to c} \; 
%{\sst +0.25\atop\sst -0.26}_{b\to u}}^{\pm 15\%}
%\ea
%Notice that the quoted theory error (3\%) is much smaller than experimental
%errors (15\%), dominated by the uncertainty in the residual contribution
%from $b\to c$ decays, and by the uncertainty in the modeling of the
%final state in $b\to u$ decays.

 NLO corrections to the extraction of $\vub$ using low-mass $X_u$
have recently been calculated~\cite{DeFazio:1999sv}. An  estimate of
the residual systematic uncertainty in the range of $\sim 10\%$ was
quoted.

It is important to point out that the accuracy of future data from the
B factories will be dominated by exclusive decays. For example, the
studies presented in the BaBar Physics Book~\cite{Harrison:1998yr} anticipate
$\delta_{exp}^{excl} \sim 2.5\%$, with $\delta_{exp}^{incl} \sim
15\%$.
One can presumably conclude that efforts for an improved theoretical
determination of $\vub$ should concentrate on the calculation of
exclusive form factors, where lattice QCD can play a major role.

\subsection{$\bf \vert \vtd\vert $}
The remaining measurements/constraints on \vckm\ require loop-level
processes, since they all involve couplings of the top quark to the
$s$ and $d$ quarks, none of which can be measured today directly in
top decays.  To interpret these measurements, one must therefore
assume the SM. Massive BSM particles could in fact propagate within
the loops, and spoil the connection to the \vckm\ entries.  Within the
SM we have $V_{ts}=-V_{cb}$, and the last independent entry is
therefore $\vert V_{td} \vert$. Its { \em less indirect} measurement
comes from the mixing of $B^0-\overline{B}^0$, mediated by box
diagrams similar to those occurring in $K^0-\overline{K}^0$ mixing:
\be
\Delta m_{B_d} \propto 
\vert V_{td} V_{tb}^* \vert^2 \; f_{B_d}^2 \, B_{B_d}
\ee
Examples of BSM contributions to this process are given, for example,
by SUSY boxes, where stop and charginos can propagate in the loop
replacing top and $W$'s.

The current value of  $\vert V_{td} \vert$, as obtained from the
official average of the LEP, SLD and CDF results, is given by~\cite{LEPBosc}:
\ba
   \Delta m_{B_d} &=& 0.481\pm 0.017~\mbox{ps}^{-1} \\ 
   \vert V_{td} \vert &=&  (8.4\pm 1.6)\times 10^{-3}
\ea
$\vert \vtd \vert$ gives an independent constraint on $(\rho,\eta)$,
represented by a circle centered around the point $\rho=1$.
%\centerline{\epsfig{figure=vtd.eps,height=0.2\textheight,clip=}}
Perfect knowledge of $\vert V_{td} \vert$ and of $\vert V_{ub} \vert$,
therefore, would give unambiguous evidence of \cpviol\ in the CKM
model, in spite of the fact that neither of the two observables is by
itself CP violating.

In practice, the extraction of $\vert V_{td} \vert$ from $\Delta
m_{B_d}$ is limited theoretically by the imprecise~\cite{Sharpe} knowledge of 
\be 
        f_{B_d} \, \sqrt{B_{B_d}} \; = \; 210 \pm 40~\mbox{MeV}
\ee 
Notice that the experimental accuracy is at the level of 3\%, and will
significantly improve over the next few years, using the results from
the Tevatron and the B factories.  Will lattice calculations ever
match this accuracy?  Progress will hopefully come form the use of the
experimentally determined values of $ f_{D_s}$ and $ f_{D}$ (extracted
from $D\to \ell \nu$ decays), together with lattice QCD estimates of $
f_{D}/ f_{B_d}$ and $ f_{D_s}/ f_{D}$.  One hopes to reach an accuracy
of 5\% on $B_B$ (lattice QCD) and of 5\% on $f_{B_d}$, with a total
uncertainty of 5--10 MeV on $f_{B_d} \, \sqrt{B_{B_d}}$. The impact of
these improvements is discussed in~\cite{Parodi:1999nr}.

\subsection{\bf $\bf \Delta m_s$ and $\bf \vert \vts / \vtd \vert$ }
A large fraction of the theoretical uncertainties on $\vert\vtd \vert$ 
disappears in the ratio:
\be
 \frac{\Delta m_s}{\Delta m_d} \; = \; \frac{m_{B_s}}{m_{B_d}} \;
  \left \vert \frac{V_{ts}}{V_{td}} \right\vert^2 \xi^2 \ee
with
\be
  \xi \; = \; \frac{f_{B_s} \, \sqrt{B_{B_s}}}{f_{B_d} \,\sqrt{ B_{B_d}}}
  \; = \; 
           1.11 \pm 0.13
\ee
with the error dominated by $f_{B_s}/f_{B_d}$, since 
$B_{B_s}/B_{B_d}=1.01\pm0.01$~\cite{Gimenez}.

$B_s$ oscillations have not been observed as yet.
The current 95\%~CL limit from LEP/SLD is~\cite{LEPBosc}:
\be
 \Delta m_s \; > \; 14.3~\mbox{ps}^{-1} \quad
(x_s\gsim 21)
\ee
leading to:
\be
\vert V_{td} \vert / \vert V_{ts} \vert \; < \; 0.24 \Rightarrow
\vert V_{td} \vert <0.010
\ee
At small $\rho$, this constraint is as strong as the
measurement of $\vert V_{td} \vert$!

\subsection{Global fits in the $(\rho,\eta)$ plane}
\begin{figure}[htb]
\includegraphics*[width=0.45\textwidth]{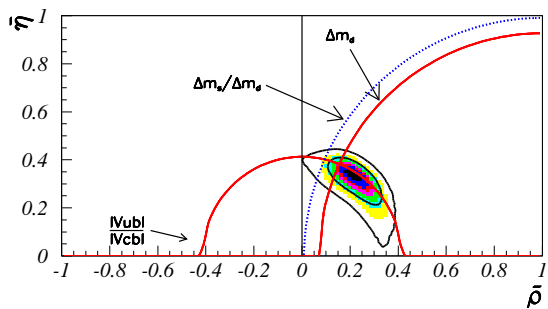}
\vspace*{-1cm}
\caption{Confidence-level countour plots, form global fits to the
$\rho$ and $\eta$ parameters using $B$-decay observables (from
ref.~\cite{Parodi:1999nr}).}
\label{fig:PRS}
\end{figure}

The set of measurements discussed above has been analysed recently by
several groups~\cite{Parodi:1999nr},
\cite{Mele:1998bf}-\cite{Ciuchini:1999} which find evidence, at better
than 95\%CL, for $\eta > 0$.  I recall here the results of the study
by Parodi et al. (PRS)~\cite{Parodi:1999nr}.  The countour lines of
their fits are shown in fig.~\ref{fig:PRS}.
Evidence for $\eta>0$ is even stronger if information from the
$\epsilon$ parameter in $K^0\bar K^0$ mixing is used
(see~\cite{Parodi:1999nr}).
%Inclusion of the constraints from the \cpviol\ parameter in the $K$
%system $\vert \epsilon_K\vert$ gives: \\
%\centerline{\epsfig{figure=parodi_eps.eps,height=0.33\textheight,clip=}}

Removing the various inputs for the fit one at the time, PRS obtain
the results shown in table~\ref{tab:PRS}
\begin{table*}[htb]
\begin{center}
\vspace*{-1cm}
\caption{Values of the parameters and inputs used in the fits to the
$B$-decay observables described above, in addition to
$\epsilon_K$. Results and table taken from
ref.~\cite{Parodi:1999nr}. The two values given for $\vub/\vcb$
correspond to the CLEO and LEP detemrinations, respectively.}
\label{tab:PRS}
\begin{tabular}{@{}ccc}
\hline
Parameter     & Fitted value
& Present value \\ \hline 
$\Delta m_s$  &  $(14.8 \pm 2.8)~ps^{-1}$  & $>$ 12.3 ps$^{-1} 
                                          ~{\rm at}~95\%{\rm CL}$ \\
$\vubovcb$  & $ 0.097{+0.033 \atop -0.022}$ & $0.080\pm 0.017/0.104\pm0.019$ \\
$B_K$ &  $ 0.87{+0.34 \atop -0.20}$ &  $ 0.86 \pm 0.09$ \\
{ $f_{B_d}\sqrt{B_{B_d}}$} & {  ($233\pm 13$) MeV}
  & ({ $210{+39\atop -32}$) MeV} \\
$\overline{m_t}(m_t)$  &  $(179{52\atop -34})$ GeV  & $(167 \pm 5)$ GeV  \\
$\vert V_{cb}\vert$   & $ (42{+8.0\atop -4.0})\times 10^{-3}$ & 
                     $ (40.0\pm2.2 )\times 10^{-3}$ \\
{ $\sin2\beta$} & { \bf $0.725{+0.050\atop -0.060}$} &
{ \bf $0.79{+0.41\atop-0.44}$}
 { (CDF~\cite{Affolder:1999gg})}\\
$\sin2\alpha$ & $-0.26{+0.29\atop-0.28}$ & --- \\
$\gamma$ & $(59.5{+8.5\atop-7.5})^{\circ} $ & ---  \\ \hline
\end{tabular}
\end{center}
\end{table*} 

Notice that:
\begin{itemize}
\item $B_K>0.60$ at 98.4\%~CL
\item The current accuracy on  $\sin2\beta$, obtained from the overall
fit of non CP-violating observables, is similar to that expected after
$\sim$ 3-4 yrs of running at a B-factory! The fitted value of
$\sin2\beta$ is consistent at this time with the CDF
measurement~\cite{Affolder:1999gg}.
\item $f_{B_d}\sqrt{B_{B_d}}$, as determined from the global fit, 
has a better accuracy than what can be achieved today from lattice
QCD! 
\item The best fit value for $x_s$ suggests that the observation of
$B_s$ mixing is behind the corner. While LEP and SLD may not be able
to see it, having almost completed their data analysis, CDF should
have no problem to detect it during its forthcoming new run.
\end{itemize}

In view of the above results, and of the estimates for the performance
expected from the upcoming experiments, it is interesting to ask
how far can the theoretical accuracy be pushed, and what would one
gain from this improved accuracy. Table~\ref{tab:future} contains a
summary  of the theoretical uncertainties, and exptected experimental
ones (1--3 yrs of B-factory, from the BaBar book~\cite{Harrison:1998yr}).
\begin{table*}[htb]
\begin{center}
\caption{Current theoretical uncertainties, and projected experimental
uncertainties, for $3^{rd}$ generation entries of the CKM matrix.} 
\label{tab:future}
\begin{tabular}{@{}lcll}
\hline
 Quantity &  Theory unc. & Main source & Projected experimental unc.\\ \hline
 $\vert \vcb^{\rm incl} \vert$ &  $\pm 5\%$ &   $m_Q$, PT, ${\cal
  O}(1/m_Q^3)$
    &  $<$1\%  \\
 $\vert \vcb^{\rm excl} \vert$ &  $\pm 5\%$ &   $F_{D^*}(1):~{\cal
  O}(1/m_Q^3),D^{**}$ &  1.5\%\\
 $\vert \vub^{\rm incl} \vert$ &  $\pm (4-10)\%$ &  
$m_b$, PT &  $\sim 15\%$\\
 $\vert \vub^{\rm excl} \vert$ &  $\pm 15\%$ &  FF's &
      $\sim 2.5\%$ \\
 $\vert (\vub/\vcb)^{incl} \vert$ &  $<3\%$ &  $m_Q$ &
      ? \\
 $\vert \vtd \vert$ & $\pm 20\%$ &  $f_{B_d}\sqrt{B_B}$ &
      $\delta(\Delta m_d)\sim 1\%_{stat}\oplus1.5\%_{\Gamma^0}$ \\
 $\vert \vts/\vtd \vert$ &  $\pm 5\%$? &   $f_{B_d}/f_{B_s}$ & 
      CDF (Run II): $\delta(\Delta m_s)\sim 2\%$ \\
\hline
\end{tabular}
\end{center}
\end{table*}

%Possible improvements from use of $f_{D_s}$ extracted from $D_s\to \mu^+\nu$:
%\[  f_{D_s}^{world}=241\pm32 \quad {\rm vs} \quad
%    f_{D_s}^{latt}=220\pm10_{stat}\pm20_{syst}~{\rm MeV} \]
%together with
%  lattice evaluations of $f_{B_d}/f_{D_s}$, 
%\[
%     \frac{f_{B_d}}{f_{D_s}} \=
%\begin{array}{ll}
%           0.78 \pm 0.04 & \mbox{\darkviolet FNAL} \\
%           0.75 \pm 0.03 {+0.07\atop -0.00}_{\rm unquen} 
%                                & \mbox{\darkviolet MILC} \\
 %          0.71 \pm 0.04 {+0.07\atop -0.00}_{\rm unquen} 
 %                               & \mbox{\darkviolet APE}
 %          \end{array}
%\]
%to extract $f_{B_d}$:
%\[
%    f_{B_d}=181\pm24_{exp}\pm7_{\rm th stat}{+20\atop -5}_{\rm th
%    syst}
%\]

 Some comments:
 \begin{itemize}
\item Once $\Delta m_s$ will be measured, the ratio $\Delta m_s / \Delta m_d$
will provide the best determination of $\vert 1-(\rho+i\eta)\vert$ (with
the SM assumption)
\item This determination could also be less affected by possible
  contaminations from new physics. For example, MSSM $(\tilde t,\tilde
  \chi^\pm)$ boxes would cancel in the ratio
\item $\Delta m_d$ can then be used to explore possible deviations
  from the SM. Assuming the SM relation $\vert \vtd\vert = \vert
  \vtd\vcb/\vts \vert$, the reach for these explorations will be by
  and large determined by the uncertainty in $f_{B_d}\sqrt{B_B}$,
  until this reaches an accuracy comparable to $\vert \vcb \vert$.
\end{itemize}

As an example of the impact of improved accuracies in
$f_{B_d}\sqrt{B_B}$ for searches of new physics effects, 
consider the constraints on
$m(\tilde t),\;m(\tilde\chi^\pm)$ from $B_d$ mixing.
For a higgsino-like chargino,
the supersymmetric box contribution to $B_d$ mixing~\cite{Brignole:1996js}
is given by:
\be
\frac{(\Delta m_d)_{SUSY}}{(\Delta m_d)_{SM}} \esim
   \left(\frac{m_t}{m_{\tilde\chi}}\right)^2 \frac{1}{2.2\sin^4\beta}
   G(\frac{m_{\tilde t}}{m_{\tilde\chi}}).
\ee
The function $G(x)$ is a slowly varying function of $x$, given
explicitly in ref.~\cite{Brignole:1996js}, which takes the value
$G(1)=\frac{1}{3}$.
In the most conservative case of large $\tan\beta$ ($\sin\beta=1$), we
get the following effects:
\begin{center}
\begin{tabular}{@{}lcccc}
\hline
$m_{\tilde\chi^\pm}=m_{\tilde t}$ (GeV) & 100 & 150 & 200 & 300 \\
$(\Delta m_d)_{SS}/(\Delta m_d)_{SM}$ & 0.46 & 0.21 & 0.12 & 0.05 \\
\hline
\end{tabular}
\end{center}
No effect would be seen in the ratio $\Delta m_s / \Delta m_d$, which
could then be used to set the value of the SM expectation for $\vert
\vtd\vert$.
A 5\% accuracy on $f_{B_d}\sqrt{B_B}$ would then probe a region of SUSY
masses well beyond the reach of LEP2 and presumably Tevatron.

The range of masses $m(\tilde t)\lsim m(t)$ is particularly
interesting for another phenomenon which has connections to lattice
calculations: this is  baryogenesis at the EWK phase transition, a
subject which was reviewed in this Conference by Fodor~\cite{Fodor}.

A second interesting example comes from tests of 
different speculations on the underlying structure of the quark mass
matrix. Several models of quark mass matrices~\cite{Fritzsch:1979zq}
predict for example the following relations:
\be
 \label{eq:barb1}
\left\vert \frac{\vub}{\vcb} \right\vert = \sqrt{\frac{m_u}{m_c}}
\quad
 \left\vert \frac{\vtd}{\vts} \right\vert = \sqrt{\frac{m_d}{m_s}}
\ee
Relations of this type emerge naturally, for example,
in hierarchical $U(2)$ models for the flavour symmetry of
quarks and leptons~\cite{Pomarol:1996xc}.

\begin{figure}
\begin{center}
\includegraphics*[width=0.45\textwidth]{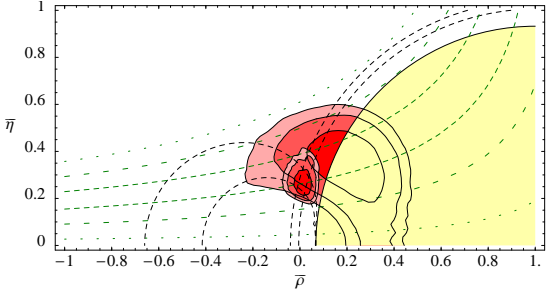}
\vspace*{-1cm}
\caption{\label{fig:barbieri} Confidence-level profiles for
$\rho+i\eta$~\cite{Barbieri:1998qs}. The smaller regions incorporate
the constraints given by the quark-mass relations in
eqs.~(\ref{eq:barb1}).}
\end{center}
\end{figure}

These relations can be used to predict $\rho,\eta$, or in turn can be
tested against the extractions of $\rho,\eta$ from the data.
Using as inputs:
\ba
     Q&\equiv& \frac{m_s/m_d}{\sqrt{1-(m_u/m_d)^2}} =22.7\pm0.8 \\
     m_u/m_d &=& 0.553\pm 0.043 \\
     m_c/m_s &=& 8.23\pm 1.5
\ea
Barbieri, Hall and Romanino~\cite{Barbieri:1998qs}
obtained the  results given in fig.~\ref{fig:barbieri}  for $\rho+i\eta$.
This determination of $\rho+i\eta$ is currently more precise than the
SM fits. Improvements in the SM fits, and in the determination of the
light quark masses, will allow stringent tests of these texture
scenarios!

\section{$\epsilon'/\epsilon$}
Two new results have appeared 
in 1999, from analyses of data subsets from  KTeV~\cite{Alavi-Harati:1999xp}
(FNAL) and NA48~\cite{Fanti:1999nm} (CERN):
\ba
&&{\rm Re}\frac{\epsilon'}{\epsilon} = \nn \\
&& \begin{array}{ll}
 (28.0\pm 2.8_{syst} \pm 3.0_{stat} )\cdot 10^{-4} & {\rm KTeV }\\
 (18.5\pm 4.5_{syst} \pm 5.8_{stat})\cdot 10^{-4} & {\rm NA48 }\\
 (21.2\pm 2.8)\cdot 10^{-4} & \langle{\rm World}\rangle
   \end{array}
\nn
\ea
These values of \epspoeps\ firmly establish the existence of direct
CP violation in the Kaon system.
These results correspond to a small fraction of the amount of data
that will be analysed in the coming years:
\begin{itemize}
\item KTeV analysed only 25\% of total data. 
\item NA48 analysed only 25\% of available data. More data
  have been taken in 1999, and yet more will be collected in the 
 year 2000. The final sample size will be approximately 10 times
larger than the presently analysed one.
\item KLOE at the Frascati $\Phi$-factory has recently started taking
data~\cite{Kloe}.  They expect to reach
  $\delta_{stat}\sim 10^{-3}$ within 1999, and to start the dive
towards the $10^{-4}$ sensitivity level in the close future.
\item  In all cases the leading contributions to
  $\delta_{syst}$ scale like $\delta_{stat}$. 
\item  An accuracy of
  $\calO( 10^{-4})$ is therefore expected overall within 2-3 years.
\end{itemize}
What about the theoretical predictions?

\subsection{Crude overview of $(\epsilon'/\epsilon)_{theory}$ }
The following discussion is based on the recent good reviews in  
refs.~\cite{Bosch:1999wr} and~\cite{Buras:1999tb,Bertolini:1998vd}.

\begin{figure}
\begin{center}
\includegraphics*[width=0.45\textwidth]{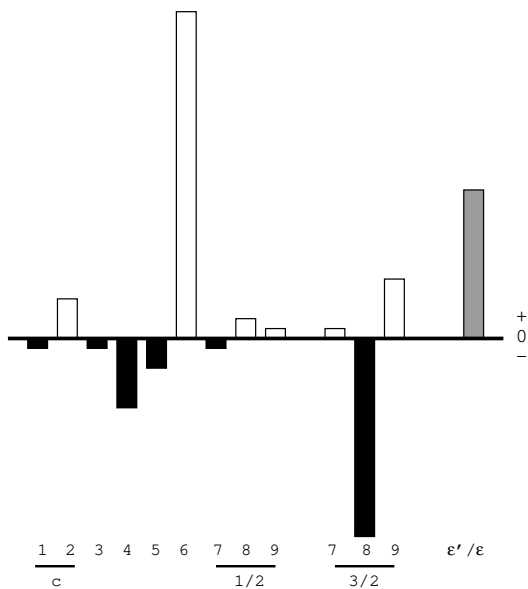}
\vspace*{-1cm}
\caption{\label{fig:ciuchini} Breakdown of the contributions from all
  possible $\Delta S=1$ 4-fermion operators, as estimated by the
  Rome group~\cite{Ciuchini:1995cd}.}
\end{center}
\end{figure}

\begin{figure}
\begin{center}
\includegraphics*[width=0.45\textwidth]{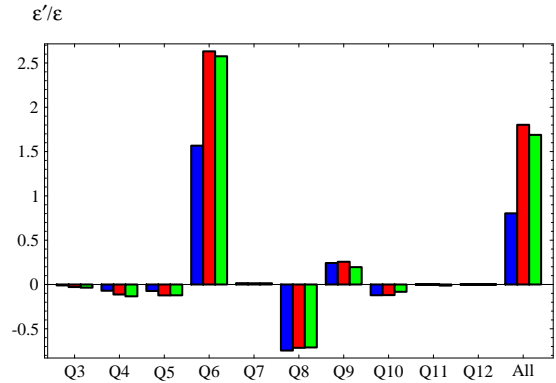}
\vspace*{-1cm}
\caption{\label{fig:bertolini}
  Breakdown of the contributions from all possible $\Delta S=1$
  4-fermion operators, as estimated by the Trieste group in the
  $\chi$QM~\cite{Bertolini:1998vd}. LO (black columns); 
   chiral 1-loop (grey columns); 
   full ${\cal O}(p^4)$ (light-grey).}
\end{center}
\end{figure}

To first approximation \epspoeps\ is dominated by the contribution of
two operators:
\ba
{\rm Re} {\epspoeps} = 13 \; {\rm Im}\lambda_t
  \,\left[\frac{130~\rm{MeV}}{m_s(m_c)}\right]^2 \;
  \left(\frac{\lambdamsb}{340~{\rm MeV}} \right) \times && \nn \\
\left[B_6^{(1/2)}(1-\Omega_{\eta+\eta'})-0.4\,B_8^{(3/2)}
  \left(\frac{m_t}{165~{\rm GeV}}\right)^{2.5} \right] &&
\ea
where:
\begin{itemize}
\item $B_{6,8}$ are the bag parameters of the QCD and EWK penguin
  operators:
{\small 
\ba
Q_6&=&(\bar s_\alpha \, d_\beta)_{V-A}\; \sum_{q=u,d,s}\,(\bar q_\beta
q_\alpha)_{V+A} \\
Q_8&=&(\bar s_\alpha \, d_\beta)_{V-A}\; \sum_{q=u,d,s}\,e_q(\bar q_\beta
q_\alpha)_{V+A} 
\ea
}
\item Im$\lambda_t={\rm Im}[V_{td}\,V_{ts}^*]=A^2\,\lambda^5\,\eta$
\item $\Omega_{\eta+\eta'}\sim 0.25$ is an $SU(3)$ breaking parameter.
\end{itemize}
The contribution of the other $\Delta S=1$ operators is strongly
suppressed, as can be seen from figs.~\ref{fig:ciuchini}
and~\ref{fig:bertolini}.  Because of the possible large cancellation
between the contributions of the two operators, the choice of input
parameters is critical to determine the value and the uncertainty of
the theoretical prediction for \epspoeps.


In particular, the value of the non-perturbative matrix elements for
the $Q_{6,8}$ operators is critical. Different approaches
(lattice~\cite{Gupta:1997bh}, $1/N$ expansion~\cite{Hambye:1998sm} and
chiral quark model ($\chi$QM, \cite{Bertolini:1998vd}) give results
for $B_8^{(3/2)}$ consistent with the value of $0.8\pm 0.2$.  The
situation for $B_6$ is much less clear.  Pre-1998 lattice
estimates~\cite{Kilcup:1991dj} led to values in the range $0.9\pm
0.3$. $\chi$QM calculations~\cite{Bertolini:1998vd} predict
$B_6^{(1/2)}=1.4\pm0.4$, with a strong dependence on the value of the
strange-quark mass. The $1/N$ expansion~\cite{Hambye:1998sm} gives
$B_6^{(1/2)}=1.6\pm0.1$, where large corrections due to $\calO(p^2/N)$
subleading terms are included.  More recent lattice estimates, based
on staggered fermions~\cite{Pekurovsky:1998jd} or domain wall
fermions~\cite{Blum:1999ib}, give negative values for $B_6$. In my
modest opinion, these results just confirm the difficulty of the
determination of $B_6$, and suggest that a rather
large uncertainty should be attached to the input value used in the
determination of \epspoeps.

In ref.~\cite{Bosch:1999wr}, the following parameter ranges have been
proposed:
\begin{itemize}
\item Im$\lambda={ (1.33\pm0.14)\cdot10^{-4}}$, from CKM fit.
\item ${m_s}^{(m_c)}= 130\pm25$~MeV, 
  mostly from Lattice, QCDSR. 
\item \lambdamsb=$ 340\pm50$~{ MeV}, 
from $\as(M_Z)=0.1185\pm0.003$. 
\item $B_8= 0.8\pm0.2$
\item $B_6= 1\pm0.3$
\end{itemize}
Normalizing to the central values for the input parameters,
and neglecting the $m_t$ dependence, one gets approximately:
\ba
{\rm Re}{\epspoeps} = 7\cdot 10^{-4} 
 \times  \left[1.9 \,B_6^{(1/2)}-B_8^{(3/2)}\right]
  && \nn \\
\times 
 \left(\frac{\lambdamsb}{340~{\rm MeV}} \right)
  \left(\frac{{\rm Im}\lambda_t}{1.34\times10^{-4}}\right)
  \,\left[\frac{130~\rm{MeV}}{m_s(m_c)}\right]^2 \;
  &&
\ea
which, for the central choices of the parameters, gives
\epspoeps$ \sim 7\cdot 10^{-4}$, far too small to agree with  the data!

Taking a 10\% shift in \lambdamsb, Im$\lambda$ and $m_s$ (consistent
with a variations within 1$\sigma$), and using
\epspoeps$=2\cdot10^{-3}$, we get:
\be   \left[1.9 \,B_6^{(1/2)}-B_8^{(3/2)}\right] \sim 2
\ee
With a value of $B_8$ consistent with the central theoretical value,
0.8, one can solve this equation for $B_6=2$.
 Is this value really unacceptable to lattice?

The recent analysis of 
Bosch et al gives the following results, using
different approaches to the description of the systematic uncertainties:
\begin{enumerate} 
\item Gaussian errors for experimental inputs, flat for theory
  parameters (NDR):
\be
  {\epspoeps}=(7.7{\scriptstyle{ +6.0\atop-3.5}})\cdot10^{-4}
\ee
\item Range covered by the scan of parameters (NDR):
\be  {\epspoeps}=(1.05 \to 28.8)\cdot10^{-4}
\ee
\end{enumerate}
Similar results have been obtained in the recent analysis of the Rome
group~\cite{Ciuchini:1999}, which also contains an interesting
appraisal of the overall uncertainties coming from the evaluation of
the non-perturbative matrix elements, and of the correlation between
the various inputs used in the theoretical calculations.

\subsection{New physics in \epspoeps?}
Because of the large uncertainty due to the evaluation of $B_6$, it is
clearly premature to draw conclusions on the possible presence of new
physics from the measurement of \epspoeps. Nevertheless, it is worth
mentioning that several papers have already appeared, trying to
explore how much room is available for contributions beyond the SM.
In the case of supersymmetry, one needs to resort to extensions of the
minimal scenarios, in particular to models with explicit \cpviol\ 
phases in the squark mass matrices. Possible contributions to
\epspoeps\ then emerge, proportional to ${\rm
  Im}\left(M^2_{12}\right)_{LR}$, where $M^2_{12}$ is a contribution
to the scalar quark mass matrix mixing left and right scalars of the
first two generations. In these models, contributions are also
expected to the neutron electric dipole moment ({$d_n^{SUSY}
  \propto {\rm Im}\left(M^2_{qq}\right)_{LR}$}). Current experimental
constraints ($<0.94\cdot10^{-25}\;e\;{\rm cm}$) make a possible
contribution to \epspoeps\ of order $10^{-3}$ unnatural, although
still possible~\cite{Masiero:1999ub}.

Other scenarios involve anomalous $Zd\bar s$
vertices~\cite{Colangelo:1998pm}.  In this case, anomalies in the rare
Kaon decays $BR(K^+\to \pi^+\nu\bar\nu)$ and $BR(K_L\to
\pi^0\nu\bar\nu)$ should appear~\cite{Buras:1999da}, which will be
testable in the experiments planned for the near future
(KAMI at FNAL, E949 at BNL).

\section{HIGH ENERGY JETS AND QCD}
While it is clear that the contribution to flavour physics is the most
fundamental one that lattice QCD can give at this time to HEP
phenomenology, there are other areas where progress is limited by a
poor understanding of the non-perturbative phase of QCD. Among these,
we list for example:
hadronic structure functions,
power corrections to high-$Q^2$ phenomena,
definition of the top quark mass, etc.
I will illustrate with examples the relevance of some of these points.
It is not clear to me whether lattice QCD can already contribute significantly
to these problems, but I understand that progress is in sight.

\subsection{Jet production at the Tevatron}
\begin{figure}
\begin{center}
\includegraphics[width=0.5\textwidth,clip]{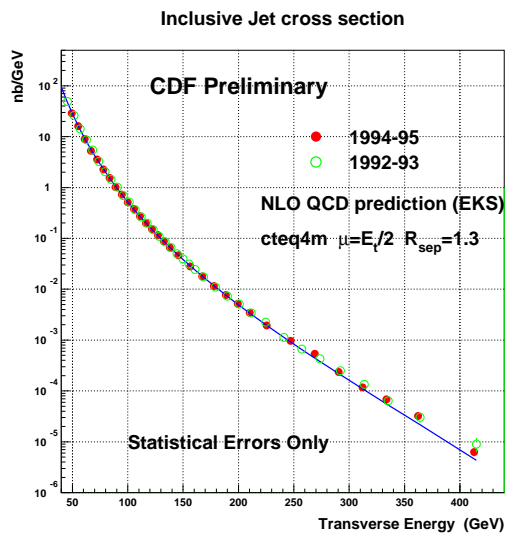}
\vspace*{-1cm}
\caption{Inclusive jet transverse energy ($\et$)
distribution as measured by CDF, compared to the absolute NLO QCD
calculation.}
 \label{fig:cdfjet} 
\end{center}
\end{figure}

\begin{figure}
\begin{center}
%\FIGURE
\includegraphics[width=0.5\textwidth,clip]{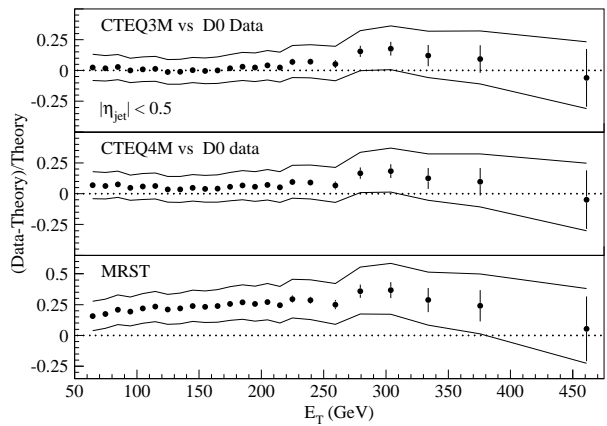}
\vspace*{-1cm}
\caption{Deviations of QCD predictions from  D0 jet
data for various sets of PDFs.}
\label{fig:d0jet_pdf}
\end{center}
\end{figure}

At the Tevatron, jets up to 450 GeV transverse momentum have been
observed~\cite{cdfjets,d0jets}. These data can be used for many
interesting purposes:
\begin{itemize}
\item Tests of QCD: calculations are available up to  NLO~\cite{jetsnlo}.
\item Extract information on the partonic densities, 
  $f_{q,g}(x,Q^2)$ at large $Q^2$.
\item Look for deviations from QCD (e.g. resonances in the dijet mass
  spectrum),  explore quark structure at small distances.
\end{itemize}
The accessible range of transverse energies corresponds to values of 
 $x \gappeq 0.5$, at $Q^2
\simeq 160,000$ GeV$^2$.  This is a domain of $x$ and $Q^2$ not
accessible to DIS experiments, such as those running at HERA.  The
current agreement between theory and data is at the level of 30 \%
over 8 orders of magnitude of cross-section, from $E_T\sim$ 20 to
$E_T\sim$ 450 GeV (see fig.~\ref{fig:cdfjet})
In spite of the general good agreement, a large dependence on the
chosen set of parton densities~\cite{Martin:1998sq,Lai:1999wy}
 is present, as shown in
fig.~\ref{fig:d0jet_pdf}. The presence of this uncertainty limits the
use of high-$\et$ jet data to set constraints on possible new physics.

An important question is therefore the following: to which extent do
independent measurements of parton densities constrain the knowledge
of PDFs at large-$x$, and what is the residual uncertainty on the jet
$\et$ distributions?

\begin{figure}
\begin{center}
%\FIGURE
\includegraphics[width=0.5\textwidth,clip]{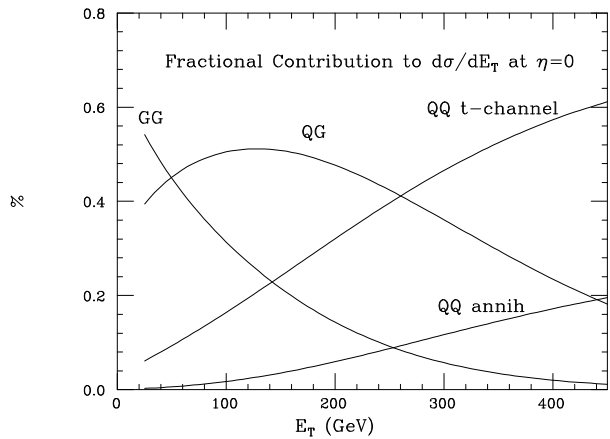}
\vspace*{-1cm}
\caption{Contributions from different initial states to the jet cross
  section at $\sqrt{s}=1.8$~TeV}
\label{fig:isfrac}
\end{center}
\end{figure}
To address this issue, let us first show what is the relative
contribution of different initial state partons to the jet cross
section. This is plotted in fig.~\ref{fig:isfrac}, where some
standard PDF set (CTEQ4M~\cite{Lai:1999wy} in this case) was chosen.
At the largest energies accessible to today's Tevatron data, 80\% of
the jets are produced by collisions involving only initial state
quarks. The remaining 20\% comes from processes where at least one
gluon was present in the initial state.

Quark densities at large-$x$ quarks are constrained by DIS data to 
within few percent, leading to an overall uncertainty on the
high-$\et$ jet rate of at most 5\%. 
What is the uncertainty on the remaining 20\% coming from gluon-induced
processes?
How are we guaranteed that the gluons are known to better than a
factor of 2, limiting the overall uncertainty to 20-30\%?

\begin{figure}
\begin{center}
%\FIGURE
\includegraphics[width=0.5\textwidth,clip]{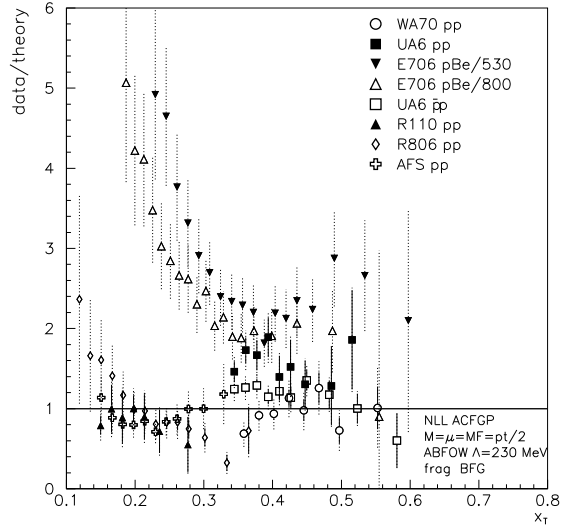}
\vspace{-1.7cm}
\caption{ Relative deviations between NLO
QCD and prompt photon data, as a function of $x_T=2\pt/\sqrt{S}$, for
various fixed target experiments.}
\label{fig:aurenche_allxs}
\end{center}
\end{figure}

\begin{figure}
\begin{center}
%\FIGURE
\includegraphics[width=0.5\textwidth,clip]{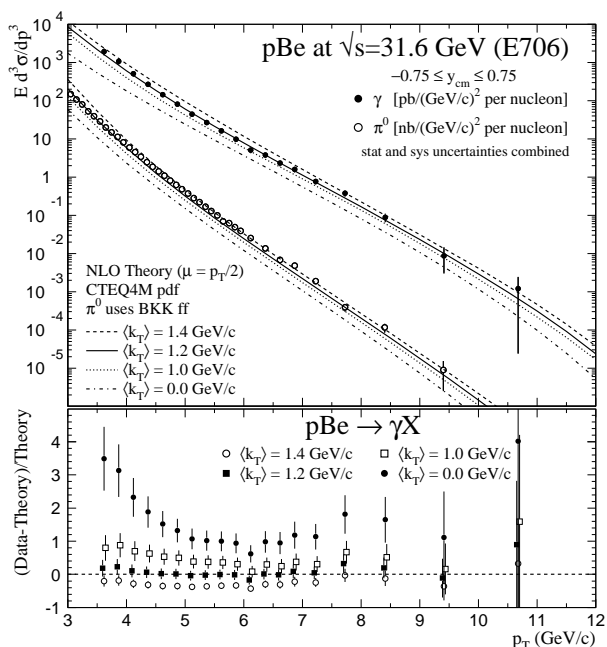}
\vspace{-1.5cm}
\caption{ Comparison of E706 data~\cite{Huston98} with NLO QCD,
before and after inclusion of an intrinsic-$k_T$.}
\label{fig:tung3}
\end{center}
\end{figure}

The only independent constraint on $ f_g(x,Q^2)$ comes from
fixed-target production of prompt photons. This process is induced at
LO by two mechanisms, $q\bar q \to g \gamma$ and $qg \to q \gamma$.
In $pN$ collisions $g(x)\gg \bar{q}(x)$, and therefore
\be
    \frac{d\sigma}{dE_T}(qg\to q\gamma) \gg
    \frac{d\sigma}{dE_T}(q\bar q\to g\gamma) 
\ee
Data from FNAL and CERN fixed target experiments can therefore be used to
extract $ f_g(x,Q^2)$ at large $x$. 
Unfortunately, a comparison of data and NLO theory shows
inconsistencies at  small
$E_T$ between the various experiments, as shown in
fig.~\ref{fig:aurenche_allxs}~\cite{Aurenche98}.
As a possible explanation for these discrepancies, 
the presence of a large
non-perturbative contribution from the intrinsic $k_T$ of partons
inside the nucleon has been suggested~\cite{Huston98,Huston99}. This gives rise
to power-like corrections to the spectrum of order $k_T/p_T$, with
possibly very large coefficients due to to the steepness of the
spectrum itself.
The effect of
the intrinsic $k_T$ is to smear the $p_T$ distribution, as shown in
fig.~\ref{fig:tung3}.
Inclusion of these effects, however, has a big impact also on the rate
at large $E_T$ (i.e. $x\sim 0.6$).  Due to the large size of the
effects, and to their intrinsic non-perturbative nature (which means
that they cannot be understood from first principles, and need to be
described by ad hoc models), it is hard to trust the theoretical
predictions obtained in this way, and to claim that prompt photons
provide a reliable way of extracting the gluon content of the proton
at large $x$. Recent theoretical improvements, such as the resummation
of large-$x_T$ logarithms
\cite{Kidonakis:1997gm,Laenen:1998qw,Catani:1998tm}, should help
understanding the large-$x$ problem, but more work is necessary.  In
conclusion, the issue of the large-$x$ behaviour of $f_g(x)$ is still
an open problem.

Concerning the possible eccess observed by CDF in its highest $\et$
jet data~\cite{cdfjets}, additional input will be available with
the data from the upcoming run of the Tevatron (due to start in the
Summer 2000), thanks to an increased energy ($\sqrt{S} \to 2$~TeV, 10\%
increase). Should the eccess be due to a problem with the gluon
density at large $x$, a discrepancy similar to the one observed at 1.8
TeV will appear at jet $\et$ values 10\% larger. If the eccess is
instead due to really new phenomena, one expects the excess to appear
at the same value of $\et$ as seen in the data at 1.8~TeV. Time will
tell!

Needless to say, any theoretical input on the expected behaviour of
the gluon density at large $x$, and on the precise way in which
non-perturbative effects such as the intrinsic motion of quarks and
gluons inside the proton are to be implemented in perturbative
calculations, would be of great interest. This is an interesting
challange to the lattice community!


\begin{thebibliography}{99}
\def    \np     #1#2#3{{Nucl. Phys.} {\bf #1} (19#2) #3}
\def    \prep   #1#2#3{{Phys. Rep.} {\bf #1}  (19#2) #3}   
\def    \pl     #1#2#3{{Phys. Lett.} {\bf #1} (19#2) #3}
\def    \plold  #1#2#3{{Phys. Lett.} {\bf #1B} (19#2) #3}
\def    \prl    #1#2#3{{Phys. Rev. Lett.} {\bf #1}  (19#2) #3}
\def    \pr     #1#2#3{{Phys. Rev.} {\bf #1}  (19#2) #3}
\def    \prd    #1#2#3{{Phys. Rev.} {\bf D#1}  (19#2) #3}
\def    \zeit   #1#2#3{{Z. Phys.} {\bf C#1}  (19#2) #3}
\def    \cmp    #1#2#3{{Comm. Math. Phys.} {\bf #1}  (19#2) #3}
\def    \ibid   #1#2#3{{\it ibid.} {\bf #1} (19#2) #3}    
\def    \hepph  #1 {{\tt hep-ph/#1}}
\def    \hepex  #1 {{\tt hep-ex/#1}}
\bibitem{PDG} C. Caso et al., Review of Particle Properties, 
Eur. Phys. J. {\bf C3} (1998), 1.
\bibitem{Neubert:1996qg}
M.~Neubert,
%``B physics and CP violation,''
Int.\ J.\ Mod.\ Phys.\ {\bf A11}, 4173 (1996)
hep-ph/9604412.
%%CITATION = IMPAE,A11,4173;%%
%\bibitem{Bigi:1997fj}
I.~Bigi, M.~Shifman and N.~Uraltsev,
%``Aspects of heavy quark theory,''
Ann.\ Rev.\ Nucl.\ Part.\ Sci.\ {\bf 47}, 591 (1997)
hep-ph/9703290.
%%CITATION = ARNUA,47,591;%%
\bibitem{LEPSC}
D. Abbaneo et al., internal note of the LEP heavy-flavour steering
group, LEPHFS 99-02,
available from http://home.cern.ch/r/roudeau/www/ \\
bc\_steering.html.
\bibitem{LEPVcb}
D. Abbaneo et al., internal note of the LEP heavy-flavour $\vcb$
working group, submitted to the 1999 EPS Conference, Tampere, 15-21
July 1999, available from http://www.cern.ch/LEPVCB/
\bibitem{Bigi:1999dv}
I.I.~Bigi,
%``Memo on extracting |V(cb)| and |V(ub)/V(cb)| from semileptonic  B decays,''
hep-ph/9907270.
%%CITATION = HEP-PH 9907270;%%
\bibitem{Hashimoto:1999yp}
S.~Hashimoto et al.,
% A.X.~El-Khadra, A.S.~Kronfeld, P.B.~Mackenzie, S.M.~Ryan and J.N.~Simone,
%``Lattice QCD calculation of anti-B --> D l anti-nu decay form factors  at zero recoil,''
hep-ph/9906376.
%%CITATION = HEP-PH 9906376;%%
\bibitem{Artuso}
M. Artuso, plenary talk at the 1999 EPS Conference, Tampere, 15-21
July 1999.
\bibitem{Hashimoto:1999bk}
S.~Hashimoto,
%``B decays on the lattice,''
hep-lat/9909136.
%%CITATION = HEP-LAT 9909136;%%
\bibitem{Behrens:1999vv}
B.H.~Behrens {\it et al.}
[CLEO Collab.],
%``Measurement of B --> rho l nu decay and |V(ub)|,''
hep-ex/9905056.
%%CITATION = HEP-EX 9905056;%%
\bibitem{Bigi:1997dn}
I.~Bigi, R.D.~Dikeman and N.~Uraltsev,
%``The hadronic recoil mass spectrum in semileptonic B decays and  extracting |V(ub)| in a model-insensitive way,''
Eur.\ Phys.\ J.\ {\bf C4}, 453 (1998)
hep-ph/9706520.
%%CITATION = EPHJA,C4,453;%%
N. Uraltsev, hep-ph/9905520.
A.H. Hoang, Z.Ligeti and A.V. Manohar, 
Phys.\ Rev.\ Lett.\ {\bf 82}, 277 (1999); 
Phys.\ Rev.\ {\bf D59}, 074017 (1999).
\bibitem{Melnikov:1998ug}
K.~Melnikov and A.~Yelkhovsky,
%``The b quark low-scale running mass from Upsilon sum rules,''
Phys.\ Rev.\ {\bf D59}, 114009 (1999)
hep-ph/9805270.
%%CITATION = PHRVA,D59,114009;%%
% (Hoang=4.20(6), Beneke/Signer=4.25(8), 
% Melnikov/Yelkhovsky=4.2(1))
%\bibitem{Hoang:1999ye}
A.H.~Hoang,
%``1S and MS-bar bottom quark masses from Upsilon sum rules,''
hep-ph/9905550.
%%CITATION = HEP-PH 9905550;%%
%\bibitem{Beneke:1999fe}
M.~Beneke and A.~Signer,
%``The bottom MS-bar quark mass from sum rules at next-to-next-to-leading  order,''
hep-ph/9906475.
%%CITATION = HEP-PH 9906475;%%
\bibitem{Gimenez:1999en}
V.~Gimenez, L.~Giusti, F.~Rapuano and G.~Martinelli,
%``NNLO unquenched calculation of the b quark mass,''
hep-lat/9909138.
%%CITATION = HEP-LAT 9909138;%%
\bibitem{LEPVub}
D. Abbaneo et al., internal note of the LEP heavy-flavour $\vub$
working group, LEPVUB 99-01, available from
http://home.cern.ch/\~battagl/vub/vub.html.
\bibitem{DeFazio:1999sv}
F.~De Fazio and M.~Neubert,
%``B --> X/u l anti-nu/l decay distributions to order alpha(s),''
JHEP {\bf 06}, 017 (1999)
hep-ph/9905351.
%%CITATION = JHEPA,9906,017;%%
\bibitem{Harrison:1998yr}
P.F.~Harrison and H.R.~Quinn
[BABAR Collab.],
``The BaBar physics book: Physics at an asymmetric B factory,''
%{\it Papers from Workshop on Physics at an Asymmetric B Factory (BaBar
%                  Collab. Meeting), Rome, Italy, 11-14 Nov 1996,
%                  Princeton, NJ, 17-20 Mar 1997, Orsay, France, 16-19 Jun
%                  1997 and Pasadena, CA, 22-24 Sep 1997}.
\bibitem{LEPBosc}
The LEP/SLD/CDF B oscillations working group, prepared for the 1999 Summer
Conferences, available through http://www.cern.ch/LEPBOSC/.
\bibitem{Sharpe} 
S.R.~Sharpe,
%``Progress in lattice gauge theory,''
hep-lat/9811006.
%%CITATION = HEP-LAT 9811006;%%
\bibitem{Parodi:1999nr}
F.~Parodi, P.~Roudeau and A.~Stocchi,
%``Constraints on the parameters of the CKM matrix by end 1998,''
hep-ex/9903063.
%%CITATION = HEP-EX 9903063;%%
\bibitem{Gimenez}
V. Gim\'enez and G. Martinelli, 
Phys.\ Lett.\ {\bf B398}, 135 (1997)
\bibitem{Mele:1998bf}
S.~Mele,
%``Indirect measurement of the vertex and angles of the unitarity  triangle,''
Phys.\ Rev.\ {\bf D59}, 113011 (1999)
hep-ph/9810333.
%%CITATION = PHRVA,D59,113011;%%
\bibitem{Ali:1999we}
A.~Ali and D.~London,
%``Profiles of the unitarity triangle and CP-violating phases in the  standard model and supersymmetric theories,''
Eur.\ Phys.\ J.\ {\bf C9}, 687 (1999)
hep-ph/9903535.
%%CITATION = EPHJA,C9,687;%%
\bibitem{Bosch:1999wr}
S.~Bosch, A.J.~Buras, M.~Gorbahn, S.~Jager, M.~Jamin, M.E.~Lautenbacher and L.~Silvestrini,
%``Standard model confronting new results for epsilon'/epsilon,''
hep-ph/9904408.
%%CITATION = HEP-PH 9904408;%%
\bibitem{Ciuchini:1999}
M.~Ciuchini, E.~Franco, L.~Giusti, V.~Lubicz and G.~Martinelli, hep-ph/9910236.
\bibitem{Affolder:1999gg}
T.~Affolder {\it et al.}
[CDF Collab.],
%``A measurement of sin 2beta from B --> J/psi K0(S) with the CDF  detector,''
hep-ex/9909003.
%%CITATION = HEP-EX 9909003;%%
\bibitem{Brignole:1996js}
A.~Brignole, F.~Feruglio and F.~Zwirner,
%``Phenomenological implications of light stop and higgsinos,''
Z.\ Phys.\ {\bf C71}, 679 (1996)
hep-ph/9601293.
%%CITATION = ZEPYA,C71,679;%%
%\bibitem{Bertolini:1991if}
S.~Bertolini, F.~Borzumati, A.~Masiero and G.~Ridolfi,
%``Effects of supergravity induced electroweak breaking on rare B decays and mixings,''
Nucl.\ Phys.\ {\bf B353}, 591 (1991).
%%CITATION = NUPHA,B353,591;%%
\bibitem{Fodor}
Z. Fodor, in these proceedings.
\bibitem{Fritzsch:1979zq}
H.~Fritzsch,
%``Quark Masses And Flavor Mixing,''
Nucl.\ Phys.\ {\bf B155}, 189 (1979).
%%CITATION = NUPHA,B155,189;%%
\bibitem{Pomarol:1996xc}
A.~Pomarol and D.~Tommasini,
%``Horizontal symmetries for the supersymmetric flavor problem,''
Nucl.\ Phys.\ {\bf B466}, 3 (1996)
hep-ph/9507462.
%%CITATION = NUPHA,B466,3;%%
%\bibitem{Barbieri:1996uv}
R.~Barbieri, G.~Dvali and L.J.~Hall,
%``Predictions From A U(2) Flavour Symmetry In Supersymmetric Theories,''
Phys.\ Lett.\ {\bf B377}, 76 (1996)
hep-ph/9512388.
%%CITATION = PHLTA,B377,76;%%
\bibitem{Barbieri:1998qs}
R.~Barbieri, L.J.~Hall and A.~Romanino,
%``Precise tests of a quark mass texture,''
Nucl.\ Phys.\ {\bf B551}, 93 (1999)
hep-ph/9812384.
%%CITATION = NUPHA,B551,93;%%
\bibitem{Alavi-Harati:1999xp}
A.~Alavi-Harati {\it et al.}
[KTeV Collab.],
%``Observation of direct CP violation in K(S,L) --> pi pi decays,''
Phys.\ Rev.\ Lett.\ {\bf 83}, 22 (1999)
hep-ex/9905060.
%%CITATION = PRLTA,83,22;%%
\bibitem{Fanti:1999nm}
V.~Fanti {\it et al.}
[NA48 Collab.],
%``A new measurement of direct CP violation in two pion decays of the  neutral kaon,''
hep-ex/9909022.
%%CITATION = HEP-EX 9909022;%%
\bibitem{Kloe} 
S. Bertolucci, [Kloe Collab.], presented
at the 1999 Lepton-Photon Symposium, 
\bibitem{Buras:1999tb}
A.J.~Buras,
%``CP violation and rare decays of K and B mesons,''
hep-ph/9905437.
%%CITATION = HEP-PH 9905437;%%
\bibitem{Bertolini:1998vd}
S.~Bertolini, M.~Fabbrichesi and J.O.~Eeg,
%``Estimating epsilon'/epsilon: A review,''
hep-ph/9802405.
%%CITATION = HEP-PH 9802405;%%
\bibitem{Ciuchini:1995cd}
M.~Ciuchini, E.~Franco, G.~Martinelli, L.~Reina and L.~Silvestrini,
%``An Upgraded analysis of epsilon-prime epsilon at the next-to-leading order,''
Z.\ Phys.\ {\bf C68}, 239 (1995)
hep-ph/9501265.
%%CITATION = ZEPYA,C68,239;%%
\bibitem{Gupta:1997bh}
R.~Gupta,
%``B-parameters of 4-fermion operators from lattice QCD,''
Nucl.\ Phys.\ Proc.\ Suppl.\ {\bf 63}, 278 (1998)
hep-lat/9710090.
%%CITATION = NUPHZ,63,278;%%
\bibitem{Hambye:1998sm}
T.~Hambye, G.O.~Kohler, E.A.~Paschos, P.H.~Soldan and W.A.~Bardeen,
%``1/N(c) corrections to the hadronic matrix elements of Q(6) and Q(8) in  K --> pi pi decays,''
Phys.\ Rev.\ {\bf D58}, 014017 (1998)
hep-ph/9802300.
%%CITATION = PHRVA,D58,014017;%%
%\bibitem{Hambye:1999yy}
T.~Hambye, G.O.~Kohler, E.A.~Paschos and P.H.~Soldan,
%``Analysis of epsilon'/epsilon in the 1/N(c) expansion,''
hep-ph/9906434.
%%CITATION = HEP-PH 9906434;%%
\bibitem{Kilcup:1991dj}
G.~Kilcup,
%``Electroweak matrix elements: 1990 update,''
Nucl.\ Phys.\ Proc.\ Suppl.\ {\bf 20}, 417 (1991).
%%CITATION = NUPHZ,20,417;%%
%\bibitem{Sharpe:1991ea}
S.R.~Sharpe,
%``Staggered weak matrix element miscellany,''
Nucl.\ Phys.\ Proc.\ Suppl.\ {\bf 20}, 429 (1991).
%%CITATION = NUPHZ,20,429;%%
%\bibitem{Pekurovsky:1997rr}
D.~Pekurovsky and G.~Kilcup,
%``Weak matrix elements: On the way to Delta(I) = 1/2 rule and  epsilon'/epsilon with staggered fermions,''
Nucl.\ Phys.\ Proc.\ Suppl.\ {\bf 63}, 293 (1998)
hep-lat/9709146.
%%CITATION = NUPHZ,63,293;%%
\bibitem{Pekurovsky:1998jd}
D.~Pekurovsky and G.~Kilcup,
%``Matrix elements relevant for Delta(I) = 1/2 rule and epsilon'/epsilon  from lattice QCD with staggered fermions,''
hep-lat/9812019.
%%CITATION = HEP-LAT 9812019;%%
\bibitem{Blum:1999ib}
T.~Blum {\it et al.},
%``A first study of epsilon'/epsilon on the lattice using domain  wall fermions,''
hep-lat/9908025.
%%CITATION = HEP-LAT 9908025;%%
\bibitem{Masiero:1999ub}
A.~Masiero and H.~Murayama,
%``Can epsilon'/epsilon be supersymmetric?,''
Phys.\ Rev.\ Lett.\ {\bf 83}, 907 (1999)
hep-ph/9903363.
%%CITATION = PRLTA,83,907;%%
%\bibitem{Barbieri:1999ax}
R.~Barbieri, R.~Contino and A.~Strumia,
%``epsilon' from supersymmetry with non universal A terms?,''
hep-ph/9908255.
%%CITATION = HEP-PH 9908255;%%
\bibitem{Colangelo:1998pm}
G.~Colangelo and G.~Isidori,
%``Supersymmetric contributions to rare kaon decays: Beyond the single  mass-insertion approximation,''
JHEP {\bf 09}, 009 (1998)
hep-ph/9808487.
%%CITATION = JHEPA,9809,009;%%
\bibitem{Buras:1999da}
A.J.~Buras, G.~Colangelo, G.~Isidori, A.~Romanino and L.~Silvestrini,
%``Connections between epsilon'/epsilon and rare kaon decays in  supersymmetry,''
hep-ph/9908371.
%%CITATION = HEP-PH 9908371;%%
%\bibitem{Colangelo:1999kr}
G.~Colangelo, G.~Isidori and J.~Portoles,
%``Supersymmetric contributions to direct CP violation in  K --> pi pi gamma decays,''
hep-ph/9908415.
%%CITATION = HEP-PH 9908415;%%
\bibitem{cdfjets}
 F. Abe et al., CDF Collab., \prl{77}{96}{438}.
\bibitem{d0jets}
 B. Abbott et al., D$\emptyset$ Collab., \hepex{9807018}.
\bibitem{jetsnlo}
%QCD CORRECTIONS TO PARTON-PARTON SCATTERING PROCESSES.
    F.\ Aversa, P.\ Chiappetta, M.\ Greco and J.Ph. Guillet,
    \np{327}{89}{105}; \\
%\bibitem{Ellis:1990ek}
S.D.~Ellis, Z.~Kunszt and D.E.~Soper,
%``The One Jet Inclusive Cross-Section At Order Alpha-S**3 Quarks And
%                  Gluons,''
Phys.\ Rev.\ Lett.\ {\bf 64}, 2121 (1990);\\
%%CITATION = PRLTA,64,2121;%%
%\bibitem{Giele:1993dj}
W.T.~Giele, E.W.~Glover and D.A.~Kosower,
%``Higher order corrections to jet cross-sections in hadron colliders,''
Nucl.\ Phys.\ {\bf B403}, 633 (1993)
hep-ph/9302225.
%%CITATION = NUPHA,B403,633;%%
\bibitem{Martin:1998sq}
  A.D.~Martin, R.G.~Roberts, W.J.~Stirling and R.S.~Thorne,
  %``Parton distributions: A New global analysis,''
  Eur. Phys. J. {\bf C4}, 463 (1998)
  hep-ph/9803445.
  %%CITATION = EPHJA,C4,463;%%
\bibitem{Lai:1999wy}
  H.L.~Lai {\it et al.}
  [CTEQ Collab.],
  hep-ph/9903282.
  %%CITATION = HEP-PH 9903282;%%
\bibitem{Aurenche98}
  P.\ Aurenche, et al., \hepph{9811382}
\bibitem{Huston98} 
  L. Apanasevich et al., E706
  Collab., \prl{81}{98}{2642}.
\bibitem{Huston99}
 L. Apanasevich et al.,  \pr{59}{99}{074007}.
\bibitem{Kidonakis:1997gm}
  N.~Kidonakis and G.~Sterman,
  %``Resummation for QCD hard scattering,''
  Nucl. Phys. {\bf B505}, 321 (1997)
  hep-ph/9705234.
  %%CITATION = NUPHA,B505,321;%%
\bibitem{Laenen:1998qw}
  E.~Laenen, G.~Oderda and G.~Sterman,
  %``Resummation of threshold corrections for single particle inclusive cross-
  %                  sections,''
  Phys. Lett. {\bf B438}, 173 (1998)
  hep-ph/9806467.
  %%CITATION = PHLTA,B438,173;%%
\bibitem{Catani:1998tm}
  S.~Catani, M.L.~Mangano and P.~Nason,
  %``Sudakov resummation for prompt photon production in hadron collisions,''
  JHEP {\bf 07}, 024 (1998)
  hep-ph/9806484;\\
  %%CITATION = JHEPA,9807,024;%%
%   \bibitem%{Catani:1999hs}
  S.~Catani, M.L.~Mangano, P.~Nason, C.~Oleari and W.~Vogelsang,
  %``Sudakov resummation effects in prompt photon hadroproduction,''
  JHEP {\bf 03}, 025 (1999)
  hep-ph/9903436.
  %%CITATION = JHEPA,9903,025;%%

\end{thebibliography}
\end{document}